\documentclass[preprint,showpacs,superscriptaddress,aps,floatfix]{revtex4}

\usepackage{graphicx}
\usepackage{rotating}
\usepackage{amsmath}

\newcommand{\br}{{\bf r}}

\newcommand{\bk}{{\bf k}}
\newcommand{\brp}{{\bf r'}}

\begin{document}

\title{Performance of one-body reduced density matrix functionals for the homogeneous electron gas}

\author{N.\,N.\,Lathiotakis} 
\affiliation{Institut f{\"u}r Theoretische Physik, Freie Universit{\"a}t Berlin, Arnimallee 14, 
D-14195 Berlin, Germany}
\affiliation{European Theoretical Spectroscopy Facility}

\author{N.\,Helbig}
\affiliation{Institut f{\"u}r Theoretische Physik, Freie Universit{\"a}t Berlin, Arnimallee 14, 
D-14195 Berlin, Germany}
\affiliation{European Theoretical Spectroscopy Facility}
\affiliation{Fritz-Haber-Institut der Max-Planck-Gesellschaft, Faradayweg 4-6, 14195 Berlin,
Germany}
\affiliation{Unit\'e de Physico-Chimie et de Physique des Materi\'eaux, Universit\'e Catholique de Louvain, B-1348 Louvain-la-Neuve, Belgium}

\author{E.\,K.\,U.~Gross}
\affiliation{Institut f{\"u}r Theoretische Physik, Freie Universit{\"a}t Berlin, Arnimallee 14, 
D-14195 Berlin, Germany}
\affiliation{European Theoretical Spectroscopy Facility}

\begin{abstract}
The subject of this study is the exchange-correlation-energy functional of reduced density 
matrix functional theory. Approximations of this functional are tested by applying them to 
the homogeneous electron gas. We find that two approximations recently proposed by
Gritsenko,  Pernal, and Baerends, J. Chem. Phys., {\bf 122}, 204102 (2005),
yield considerably better correlation energies 
and momentum distributions than previously known functionals. 
We introduce modifications to these functionals
which, by construction, reproduce the exact correlation energy of the homogeneous electron gas. 
\end{abstract}
\maketitle

\section{Introduction}
Reduced-density-matrix-functional theory (RDMFT) is one possible way
to tackle the problem of electronic correlation. 
It is based on Gilbert's theorem~\cite{gilbert}, which is an extension of the
Hohenberg-Kohn theorem to non-local external potentials. 
Gilbert's theorem guarantees that the ground-state expectation value 
of any observable of a quantum mechanical system is a unique
functional of the ground-state one-body reduced-density matrix (1-RDM).
Thus, the fundamental quantity in RDMFT is the 1-RDM instead of the 
electronic density on which DFT is built upon.  
The properties of the exact 1-RDM have been the subject 
of theoretical studies for a long time~\cite{gilbert,valone,zumbach,mueller}. 
Nevertheless, only relatively recently approximate total-energy functionals of the 1-RDM 
were used in practical 
applications~\cite{bb0,UG,csanyi,csgoe,yasuda,kollmar,kios1,kios2,gritsenko,openshell,pernal,pirin,kollmar2}.
All these approximate functionals give a satisfactory account of electronic correlations in small 
atoms and molecules at the equilibrium distance. The latest generation of 
functionals performs equally well at the molecular dissociation limit. Especially this
latest success makes RDMFT particularly appealing.

A great advantage of RDMFT, 
compared to DFT, is that the exact many-body kinetic energy is easily 
expressed in terms of the 1-RDM. More specifically, the total energy of a
many-electron system in its ground state,
characterized by
an external potential $V(\br)$, can be expressed in terms of the ground-state
1-RDM, $\gamma$ as
\begin{multline}
\label{eq:E_of_g}
E_{\rm tot}\left[  \gamma \right] = E_{\rm ion} +
\int d^3\br \int  d^3\brp\: \delta(\br - \brp ) \left[ -\frac{1}{2} \nabla_{\!\!\br}^2 \right] \gamma(\br, \brp) \\
+\int d^3\br \int  d^3\brp\: \delta(\br - \brp ) V(\br) \gamma(\br, \brp) \\
+\frac{1}{2} \int d^3\br \int  d^3\brp \frac{\gamma(\br,\br)\: \gamma(\brp, \brp)}{|\br-\brp|}
+ E_{\rm xc} \left[  \gamma \right] \,.
\end{multline}
The first term in the above expression is a constant not related to the electronic degrees 
of freedom, for example the ion-ion repulsion energy. The next three terms are,
respectively, the kinetic, the external potential and the electrostatic
energy and they are known, explicit functionals of $\gamma$. 
Finally, the last term is the exchange and correlation (xc) energy 
which contains all the remaining electronic contributions to the total energy. 
The exact form of this functional
is unknown and for practical applications needs to be approximated. 
Most of the approximate functionals for the xc energy that have been introduced 
so far 
are implicit functionals of $\gamma$. They depend explicitly on
the natural orbitals $\varphi_j$, and the corresponding occupation numbers $n_j$ 
which are defined as the eigenfunctions and eigenvalues of the 1-RDM:
\begin{equation}
\int d^3\br'\: \gamma(\br, \brp) \: \varphi_j (\brp) = n_j\: \varphi_j (\br)\,.
\end{equation}

Viewed as a functional of arbitrary $\gamma$ in an appropriately defined domain, 
the functional given in  Eq.~(\ref{eq:E_of_g}) has a minimum value at the ground-state $\gamma$.
This appropriate domain is defined through subsidiary conditions for $\gamma$ known
as $N$-representability conditions. It was shown by Coleman~\cite{coleman} that there
 are two such conditions for $\gamma$, and they concern the occupation numbers
\begin{equation}
\label{eq:conditions}
\sum_{j=1}^\infty n_j = N\,, \quad 0 \leq n_j \leq 1\,,
\end{equation}
where $N$ is the total number of electrons. These conditions guarantee that $\gamma$ corresponds 
to either a pure many-body state or an ensemble of pure states. The first 
condition can be enforced in the minimization with respect to $\gamma$ through
the Lagrange-multipliers method. In that way, the quantity to be minimized becomes
\begin{equation}
F = E_{\rm tot} - \mu \left(\sum_{j=1}^\infty n_j - N\right)\,,
\label{eq:F}
\end{equation}
where $\mu$ is the corresponding Lagrange multiplier. $\mu$ was shown to be equal to the
chemical potential, i.e. the derivative of the total energy with respect to the total number
of electrons~\cite{our_gap}. Interestingly, this allows one to exploit the discontinuity of
$\mu$ as a function of the particle number for the calculation of the fundamental 
gap of materials and is a motivation for the development of RDMFT schemes for periodic systems in
order to address questions like the semiconductor and insulator gaps. 

The second of the $N$-representability conditions~(\ref{eq:conditions}) has a dramatic consequence: 
it allows for border minima in the occupation number optimization. In other words, it
allows for the the existence of a subset of the optimal occupation numbers 
which are exactly equal to either one or zero and do not satisfy the condition 
$\partial F/ \partial n_j = 0$.  
We refer to the corresponding states as pinned states.
It is rather unlikely for the exact theory to produce pinned states for most 
systems of interest. For an occupation number being exactly equal to one, the corresponding
natural orbital would be present in all determinants of the full CI expansion
with non-zero coefficients. A situation like that has not been found for small atoms
and molecules or for the HEG, where the exact solution can be calculated.
While this is true for the exact xc functional, 
all the approximate functionals which are considered 
in this work, yield pinned states with $n_j=1$ for all systems we applied them to, except
for the two-electron systems. These pinned states are core states and,
in the exact theory, they correspond to occupation numbers which are marginally smaller 
than one. Hence, as far as the optimal $\gamma$ is concerned, the approximate result, $n_j=1$, 
is perfectly satisfactory.
The important implication, however, is that $\partial F /  \partial n_j \neq 0$ for the pinned states,
and consequently $\delta F / \delta \gamma(\br, \brp)  \neq 0$ at the optimal $\gamma$.

A number of approximate functionals for $E_{\rm xc}$, including those of interest in the present 
work, can be cast into the form
\begin{widetext}
\begin{equation}
\label{eq:Exc}
E_{\rm xc} \left[  \gamma \right] = E_{\rm xc} \left[ \{ n_j \}, \{ \varphi_j \} \right]=
-\frac{1}{2} \sum_{j,l=1}^{\infty}  \int d^3\br \int  d^3\brp \: f(n_j,n_l) \:
\frac{\varphi_j^*(\br)\: \varphi_l^*(\brp)\: \varphi_l(\br)\: \varphi_j(\brp)}{|\br-\brp|} \,,
\end{equation}
\end{widetext}
i.e. they have the form of the usual Hartree-Fock exchange modified by the function
$f(n_j,n_l)$ of the occupation numbers. The first such approximation was introduced
by M\"uller~\cite{mueller} and corresponds to the function $f(n_j,n_l) = \sqrt {n_j n_l}$.
M\"uller considered a more general exponent for the occupation number product in the 
exchange-like term and found an optimal exponent of 1/2.  By modelling 
the exchange and correlation hole, Buijse and Baerends~\cite{bb0} arrived 
at the same functional. Goedecker and Umrigar~\cite{UG} (GU) considered a 
modification by explicitly removing the self-interaction (SI) terms. They also
presented~\cite{UG} a direct minimization with respect to the natural orbitals and the
occupation numbers and found correlation energies for small atomic 
systems which are in very good agreement with the exact results.  Later however, it
was realized that the GU
functional fails to reproduce the correct dissociation limit for
small molecules~\cite{staroverov,herbert}. On the other hand, the M\"uller 
functional yields the correct dissociation limit but, in all cases, overestimates
substantially the correlation energy~\cite{staroverov,herbert}. 

In the last decade, several other functionals of the 1-RDM have been 
introduced~\cite{csgoe,yasuda,kollmar,kios1,kios2,gritsenko,openshell,pernal,pirin,kollmar2} and 
applied to atomic and molecular systems. 
Recently, Gritsenko et al.~\cite{gritsenko} proposed 
improved 1-RDM functionals based on a hierarchy of repulsive corrections 
to the M\"uller functional. In that way, they attempted to correct the 
overcorrelation of this functional. The functionals corresponding to these hierarchical corrections, 
are called BBC1, BBC2, and BBC3. For all these functionals, it is essential to
divide the natural orbitals
into strongly and weakly occupied ones. This distinction appears naturally
for finite systems since usually a subset of the orbitals corresponds to occupation numbers close to one, and the rest to occupation numbers close to zero. For the BBC1 and BBC2  functionals, the function $f(n_j,n_l)$ is:
\begin{widetext}
\begin{equation}
\mbox{BBC1:\ \ \ \  }f(n_j,n_l) = \left\{  \begin{array}{rl}
                                 -\sqrt{n_j\: n_l}\,, & j\neq l, \mbox{and\ } j,l\mbox{\ weakly occupied,} \\
                                 \sqrt{n_j\: n_l}\,, & \mbox{\ otherwise,} 
                             \end{array}
\right.
\end{equation}
\begin{equation}
\mbox{BBC2:\ \ \ \  }f(n_j,n_l) = \left\{  \begin{array}{rl}
                                 -\sqrt{n_j\: n_l}\,, & j\neq l, \mbox{and\ } j,l\mbox{\ weakly occupied,} \\
                                 n_j\: n_l\,, & j\neq l, \mbox{and\ } j,l\mbox{\ strongly occupied,} \\
                                 \sqrt{n_j\: n_l}\,, & \mbox{\ otherwise.} 
                             \end{array}
\right.
\end{equation}
\end{widetext}
Finally, in the BBC3 functional the anti-bonding orbital is treated as strongly occupied
orbital. Additionally, the self-interaction terms are removed 
as in the GU functional, except for the bonding and anti-bonding orbitals. 
Gritsenko et al.~\cite{gritsenko} applied the BBC functionals to 
diatomic molecules and showed that they give an accurate description of
these molecules at both the equilibrium distance and the dissociation limit.

There is a strong motivation for the extension and application of 1-RDM functionals
to solid-state systems. This motivation stems from the success of these functionals 
in the description of electron correlation for finite systems, as well as the difficulties
of DFT methods in describing certain materials and properties such as the band gap of
semiconductors and insulators~\cite{ortiz,filippi,terakura,sharma} or the band
width of the conduction band in Na~\cite{sodium}.

A very important prototype system, which serves as a benchmark
for the performance of approximate 1-RDM functionals when applied to
periodic systems, is the homogeneous electron gas (HEG). Furthermore, this 
system can serve as a laboratory for the development of approximate functionals, 
in a fashion similar to DFT. As a consequence of translational invariance, the natural orbitals
can be chosen as plane waves and the 
search for the ground state 1-RDM is restricted to the optimization of the momentum distribution 
$n(\bk)$, which is the 
occupation number that corresponds to the plane-wave natural orbital with wavevector $\bk$. 
An important point to note is that the self-interaction terms for the plane-wave 
natural orbitals vanish. Consequently, the GU and the M\"uller functionals are identical.
Finally, approximations that involve a special treatment of single orbitals have zero effect 
in the continuous wavevector case. Thus, the BBC3 functional is identical to BBC2.

As a consequence of the rotational invariance of the HEG the occupation numbers have the 
property $n(\bk)=n(k)$, i.e. all the plane-wave natural orbitals corresponding to the same 
absolute value $k$ are degenerate (with respect to the occupation number). This allows one
to perform unitary transformations among the degenerate plane waves leading, e.g., to 
angular-momentum eigenfunctions
\begin{equation}
\label{eq:sphwaves}
\varphi_{klm} (\br) = j_l(k r)\: Y_{lm}(\Omega) 
\end{equation}
for the natural orbitals, where $j_l(k r)$ are spherical Bessel functions and $Y_{lm}$
spherical harmonics.
 Since these functions are localized in real space it is conceivable to include 
self-interaction corrections in terms of the natural orbitals (\ref{eq:sphwaves}) for the HEG. 
To our knowledge, this possibility has not been explored so far and is also beyond the aim of the
present work.

Choosing the natural orbitals as plane waves, the 
1-RDM of the HEG can be written as
\begin{equation}
\gamma(\br, \brp) = \frac{2}{V}\sum_{\bk} n(k)\: e^{i\bk(\br-\brp)}\,, 
\end{equation}
where $V$ is the volume of the system ($V\rightarrow \infty$).
Substituting this expression in Eqs.~(\ref{eq:E_of_g}) and~(\ref{eq:Exc}) we obtain for the total energy
\begin{widetext}
\begin{equation}
E_{\rm tot} = 2 \sum_{\bk_1} \frac{\bk_1^2}{2} \: n({k_1}) - \frac{1}{V}
\sum_{\bk_1, \bk_2} f\left(n({k_1}\right), n({k_2}))\: 
\frac{4\pi}{\left| \bk_1 - \bk_2 \right|^2 +\alpha^2} \,,
\label{eq:etotheg}
\end{equation}
where $\bk_1$, $\bk_2$ are wavevector indices, and $\alpha$ is a small quantity (usually 
$\sim 10^{-8}$) included 
for numerical stability.  As in 
Hartree-Fock, the external potential energy and the electronic Coulomb repulsion energy,
i.e. the third and fourth terms in Eq.~(\ref{eq:E_of_g})
cancel exactly with the first term i.e. the ion-ion interaction. 
The quantity $F$ of Eq.~(\ref{eq:F}) per particle then becomes
\begin{multline}
\label{eq:energy}
\frac{F}{N} = \frac{3}{2k_{\rm F}^3} \int_0^\infty dk_1\: 
k_1^2\,(k_1^2 -2\mu)\: n(k_1) \\ + 
\frac{3}{4\pi k_{\rm F}^3} \int_0^\infty dk_1 \int_0^\infty dk_2 \: k_1\: k_2\:
\log \left[ \frac{\left( k_1 -k_2  \right)^2+\alpha^2}
{\left( k_1 + k_2\right)^2+\alpha^2} \right]  f(n(k_1), n(k_2)) + \mu \,,
\end{multline}
\end{widetext}
where $k_{\rm F}=(9\pi /4)^{1/3}r_s^{-1}$ is the 
Fermi-wavevector of the non-interacting HEG and $r_s$ is the radius (in atomic units)
of the sphere with volume equal to the volume per electron.

Cioslowski and Pernal~\cite{ciospernal} applied the M\"uller functional
to the HEG and calculated analytically the resulting momentum distribution 
\begin{equation}
\label{eq:kios_per}
n(k) = 512 \pi \rho \; (1+4k^2 )^{-4}\,,
\end{equation}
where  $\rho $ is the electron-density-per-spin,
$\rho= 3 (8\pi r_s^{3})^{-1}$. The corresponding total energy 
per particle is independent of the density and equal to $-1/8$ Hartree. 
It is obvious that the solution of the Eq.~(\ref{eq:kios_per}) is consistent 
with the second $N$-representability constraint of Eq.~(\ref{eq:conditions})
only for $\rho\leq (512\pi)^{-1}$, i.e. $r_s \geq 5.77$. 
In other words, the M\"uller functional gives a solution without pinned states only
for $r_s \geq 5.77$.  In addition, Cioslowski and Pernal demonstrated
that the Oxford-Lieb~\cite{oxlieb} bound is violated for $\rho \geq 1.65\times10^{-3}$,
i.e. $r_s \leq 4.167$. The solution with pinned states for $r_s < 5.77$, was 
calculated by Cs\'anyi and Arias~\cite{csanyi}. More specifically, for
$r_s < 5.77$, one gets an optimal momentum distribution $n(k)$ with 
$n(k)=1$ for $k$ below a certain value $k_p$ and fractional $n(k) <1$ for $k>k_p$. 
This behavior is in complete analogy to the case of finite systems for the M\"uller
functional. Unfortunately, it is in conflict with the fact that the 
exact momentum distribution~\cite{exactmom2,gori} is a monotonically decreasing function of $k$ and
is strictly smaller than 1, i.e. there are no pinned states. Additionally,
the exact momentum distribution is concave for $k<k_{\rm F}$, it shows a
discontinuity at $k_{\rm F}$, and for $k>k_{\rm F}$ it goes to zero asymptotically. The size of the 
discontinuity is decreasing with $r_s$.

In addition to the M\"uller functional, Cs\'anyi and Arias~\cite{csanyi} considered a similar 
functional derived from a tensor product expansion of the two-body density matrix, which they
called Corrected Hartree-Fock (CHF). Unfortunately,
CHF gives zero correlation for the HEG in the high-density limit ($r_s\rightarrow 0$), coinciding
with Hartree-Fock. In the opposite limit, it strongly overcorrelates 
giving the same results as the M\"uller functional. In the intermediate region,
including the metallic densities,
the result for the correlation energy is close to the exact but its dependence on $r_s$ is
monotonically decreasing instead of increasing. In an attempt to improve over the M\"uller functional
and CHF, Cs\'anyi, Goedecker and Arias considered an improved tensor product expansion of
the two-particle density matrix~\cite{csgoe}. The resulting functional, which is called CGA,
performs very well in the high-density regime and significantly better than the previous two
functionals in the region of metallic densities. The deviation from the exact correlation 
energy increases
with $r_s$ and at higher densities CGA coincides with CHF and the M\"uller functional.

In the present work, we apply the BBC1 and BBC2 functionals
of Gritsenko et al.~\cite{gritsenko} to the HEG and compare with previous
functionals as far as the resulting correlation energies are concerned.
We also investigate other features like the resemblance of the resulting momentum 
distribution to the exact and the state-pinning.  In order to apply the BBC1,2 
functionals to the case of the HEG through Eq.~(\ref{eq:energy}), we need to 
distinguish between strongly and weakly occupied 
orbitals. For finite systems,  Gritsenko et al.~\cite{gritsenko} chose the first $N/2$ 
natural orbitals to be strongly occupied. In complete analogy, we can use a critical 
wavevector $k_c=k_{\rm F}$ below which all states are assumed to be strongly occupied 
while above they are weakly occupied.

An additional goal of the present work is to demonstrate that the HEG can be used to develop 
functionals suitable for metallic systems.  The idea is to modify the BBC1 functional 
in such a way that it yields the exact correlation energy for the HEG. This 
is achieved in two different ways: (i) For each given density we choose $k_c$ such that 
BBC1 reproduces the exact correlation energy of the HEG at that density. We call this functional 
$k_c$-functional. (ii) We introduce a function $s(r_s)$ multiplying the xc terms of Eq.~(\ref{eq:Exc})
for two weakly occupied orbitals, keeping $k_c=k_{\rm F}$. In this way, we replace the sign 
change of the BBC1 functional with the parameter $s$. Accordingly, we call this functional $s$-functional. 

In the following section we present details of the numerical implementation as well as the results 
of applying the BBCs and the 
$k_c$- and $s$-functional
to the HEG.

\section{Numerical Implementation, results\label{sec:results}}

The minimization of the energy expression~(\ref{eq:energy}) with respect to
$n(k)$ is performed using the steepest-descent method. We choose to work in 
energy-space instead
of $k$-space, i.e. we perform the variable substitution $\epsilon=k^2/2$ and solve numerically 
the minimization problem for  $n(\epsilon)$. Working in energy-space 
rather than $k$-space improves the 
stability of the numerical treatment. The energy $\epsilon$ is discretized using a
double-logarithmic mesh centered at $\epsilon_{\rm F} = k_{\rm F}^2/2$, where
the occupation varies the most. 
The upper limit of integration is chosen such that  
the momentum-distribution function has dropped to values smaller than $10^{-6}$. 
The double integration with respect to the energy is carried out using 
an adaptive grid technique capable of treating integrable singularities
like the logarithmic singularity of the present problem. 
The values of the energy-distribution function $n(\epsilon)$ in between the mesh points,
necessary for the adaptive grid method, are obtained from an interpolation scheme. 
The $N$-representability constraint $0 \leq n(\epsilon) \leq 1$ is implemented through 
the substitution $n(\epsilon) = \sin^2 [ \pi \: \theta(\epsilon) /2 ]$ and
variation with respect to $\theta(\epsilon)$. Extra care is required to avoid $n(\epsilon)$ being
falsely pinned to 0 or 1. Indeed, if for a particular 
$\epsilon$, $n(\epsilon)$ gets very close to 1 or 0 during the variation it would stay pinned at that
point. The variation with
respect to the Lagrange-multiplier $\mu$ is implemented as an external iterative procedure, 
thus achieving convergence for each value of $\mu$. 
The correct value of $\mu$ is selected by requiring the 
momentum-distribution function to integrate to the correct number of electrons for a given 
value of $r_s$. Finally, we found that a reasonable value for the parameter $\alpha$ 
in Eqs.~(\ref{eq:etotheg}) and (\ref{eq:energy}) is $10^{-8}$, which avoids both numerical 
problems as well as the dependence of the results on $\alpha$.

\subsection{Application of the BBC functionals to the HEG}

\begin{figure}[t]
\includegraphics[width=0.9\columnwidth,clip]{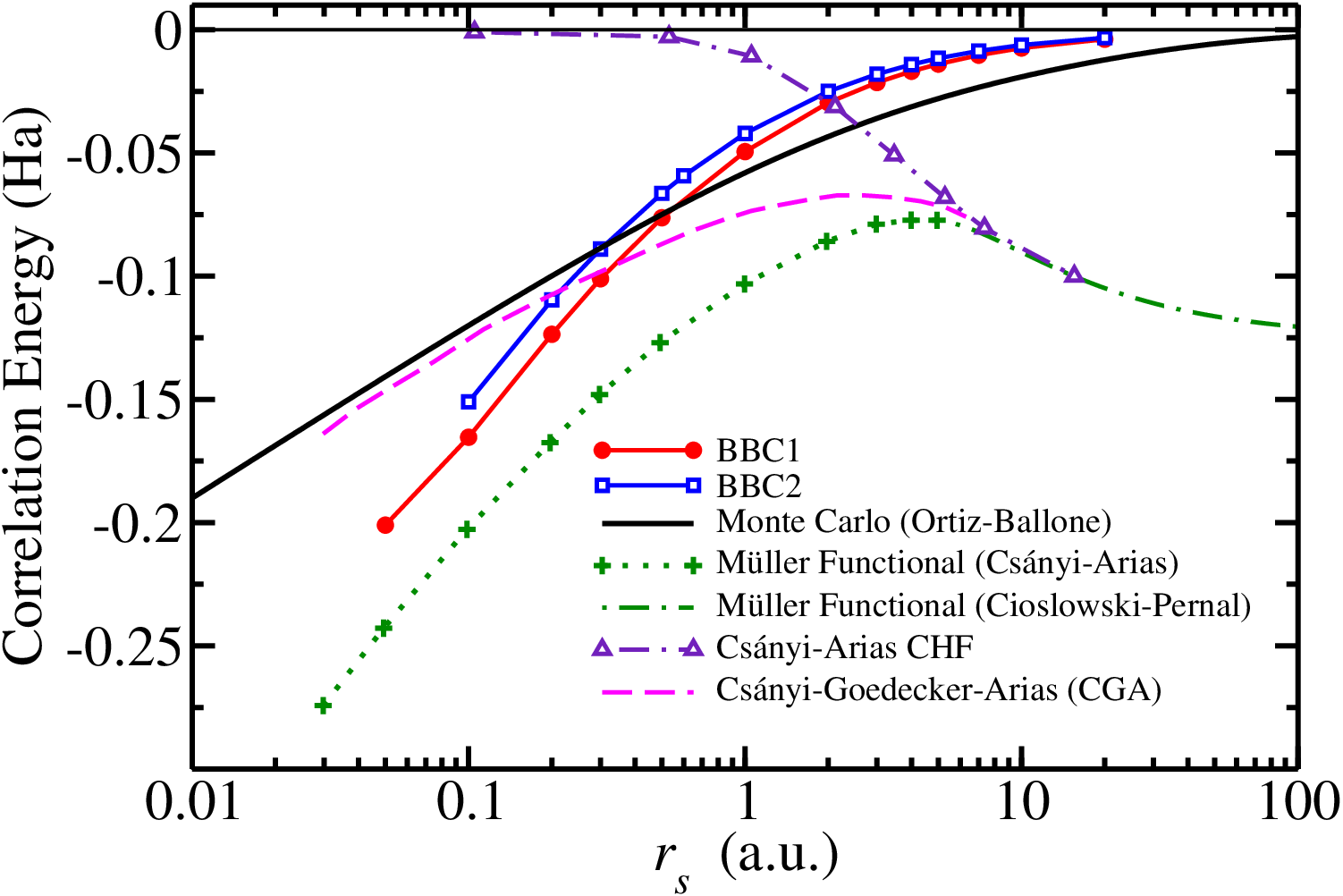}
\caption{\label{fig:correl} (Color online) The correlation energy of the HEG as a function of $r_s$
calculated with the BBC1 and BBC2 functionals compared to various other calculations.
The Monte Carlo result represents the Perdew-Wang fit~\cite{PW} of the DMC 
data of Ortiz and Ballone~\cite{exactmom2}. 
The dotted line corresponds to the numerical results by Cs\'anyi and Arias~\cite{csanyi} 
employing the M\"uller functional,
for $r_s < 5.77$. Its continuation, the dash-dotted line, for $r_s > 5.77$, stands for the
analytical results of Cioslowski and Pernal~\cite{ciospernal} employing the same functional.
The results for the CHF functional~\cite{csanyi} as well as the CGA~\cite{csgoe} are
also shown.}
\end{figure}

In Fig.~\ref{fig:correl},  we show the correlation energy of the HEG as a function of $r_s$. The
correlation energy calculated with the BBC1 and BBC2 functionals is significantly closer to
the exact than any other functional over the whole range of $r_s$.  Both functionals
also seem to reproduce the correct asymptotic limit of zero correlation for
the dilute HEG. For small densities up to metallic densities, the
BBC functionals under-correlate, i.e. the absolute value of the correlation energy is too small. 
In the dense limit, they over-correlate and the crossover is
at around $r_s$=0.5 and 0.3 for the BBC1 and BBC2, respectively. Unfortunately, in the area of low metallic
densities both functionals yield correlation energies which deviate from the 
exact values by 50\%. 
In absolute numbers, the error of BBC1 and BBC2 is of the same order as the RPA result~\cite{PW}. 
Nevertheless, in the range $0.1<r_s<1$, the BBC functionals perform remarkably well. 
Compared to all previous 1-RDM functionals, BBC1 and BBC2 offer a much better account of the 
electron correlation for the HEG. Although less accurate than the CGA in the high density 
region, they perform better for metallic densities and they reproduce the limit of zero 
correlation at the dilute HEG limit where the M\"uller functional, CHF and CGA 
fail. 

\begin{figure*}
\begin{tabular}{cc}
\includegraphics[angle=0,width=0.45\textwidth,clip]{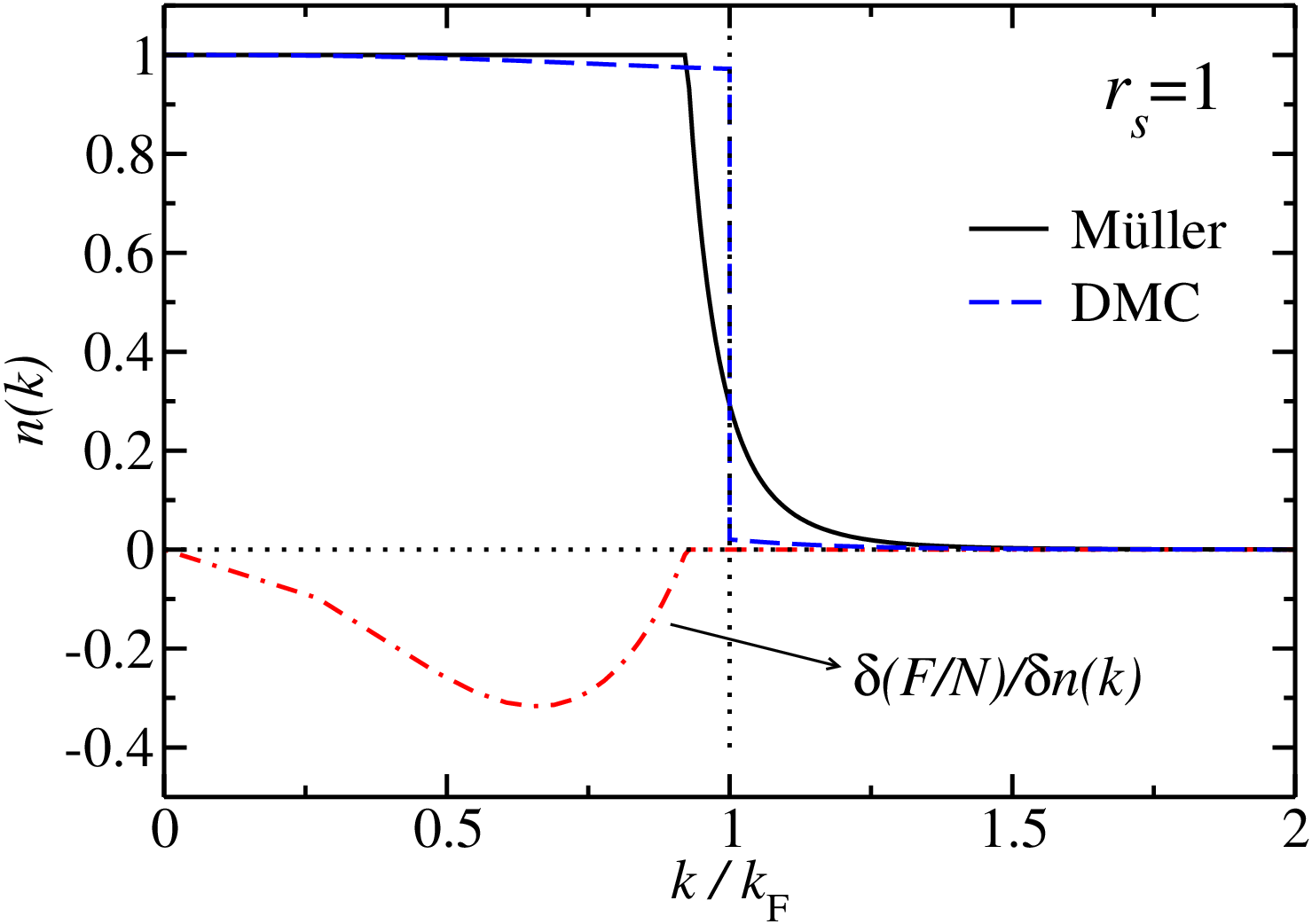}& \ \ \ \ \
\includegraphics[angle=0,width=0.45\textwidth,clip]{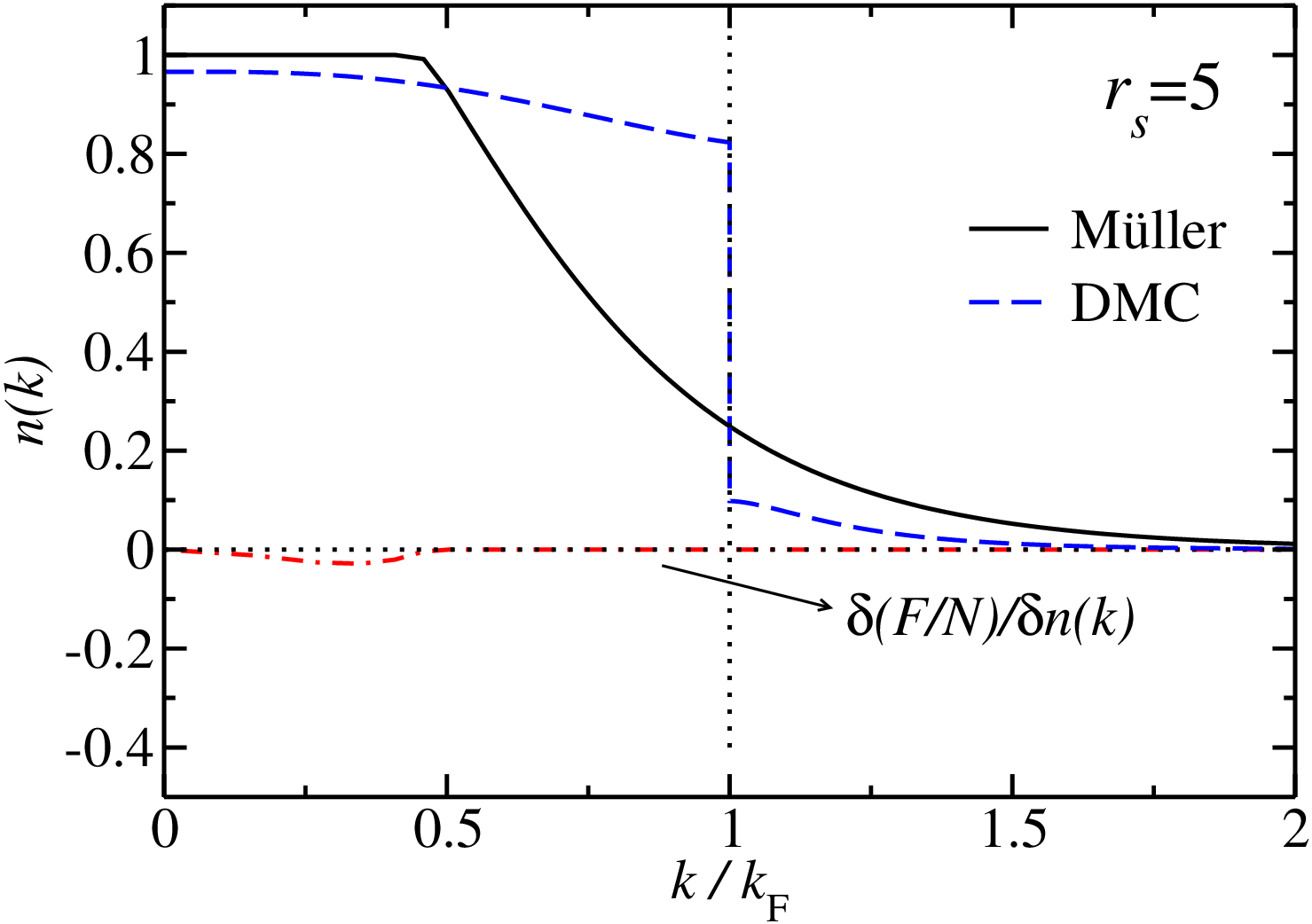}\\
\includegraphics[angle=0,width=0.45\textwidth,clip]{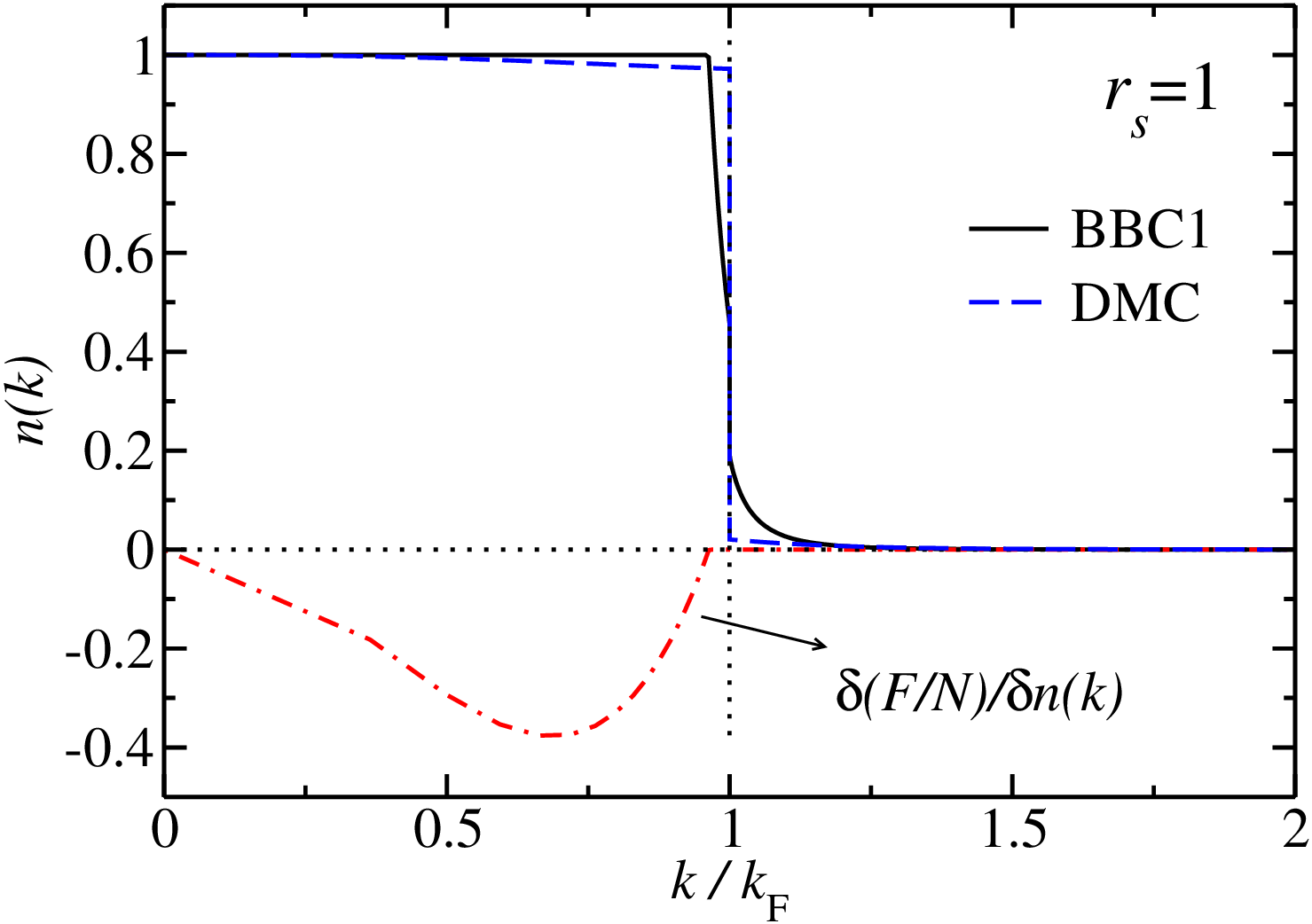}& \ \ \ \ \ 
\includegraphics[angle=0,width=0.45\textwidth,clip]{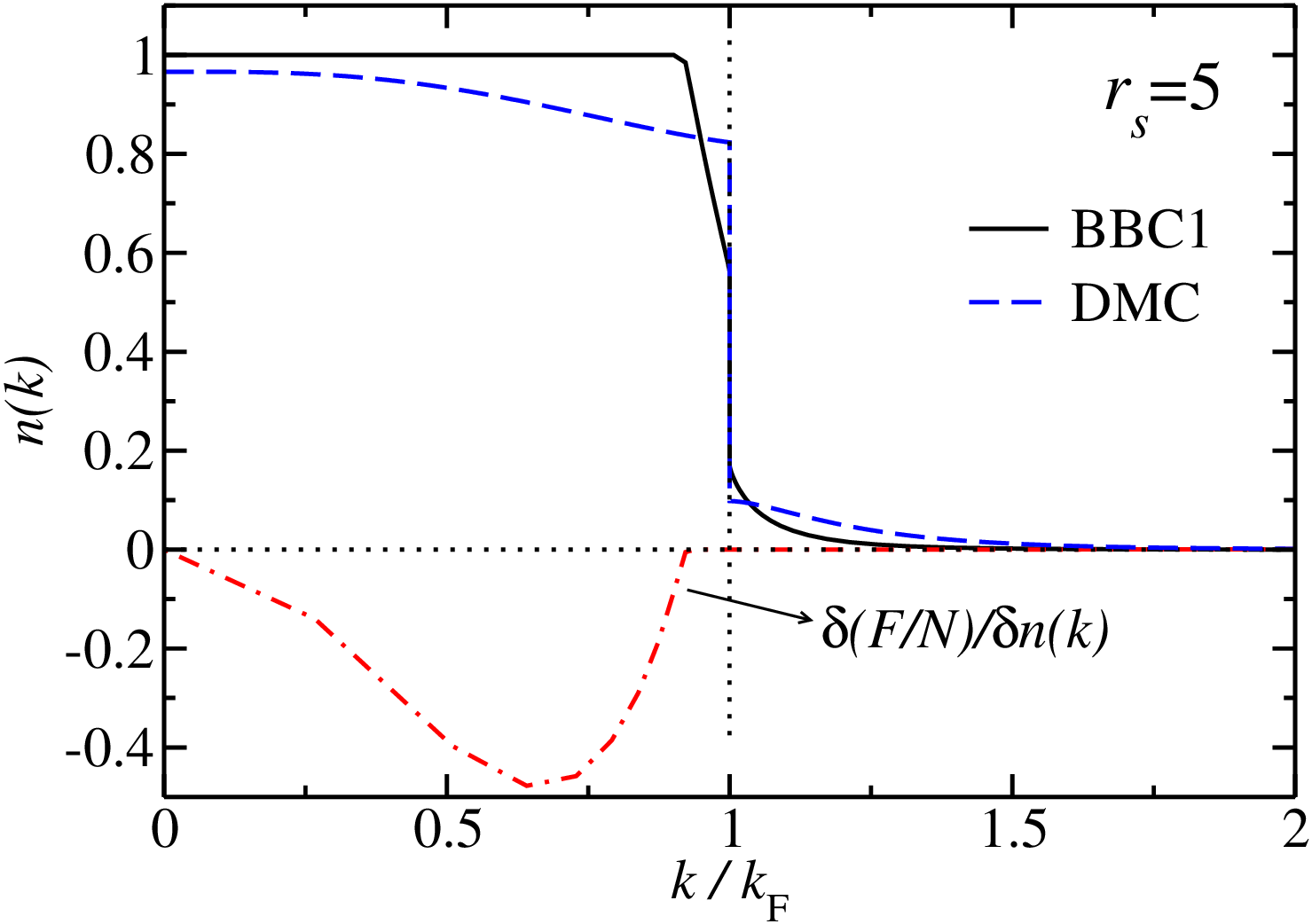}\\
\includegraphics[angle=0,width=0.45\textwidth,clip]{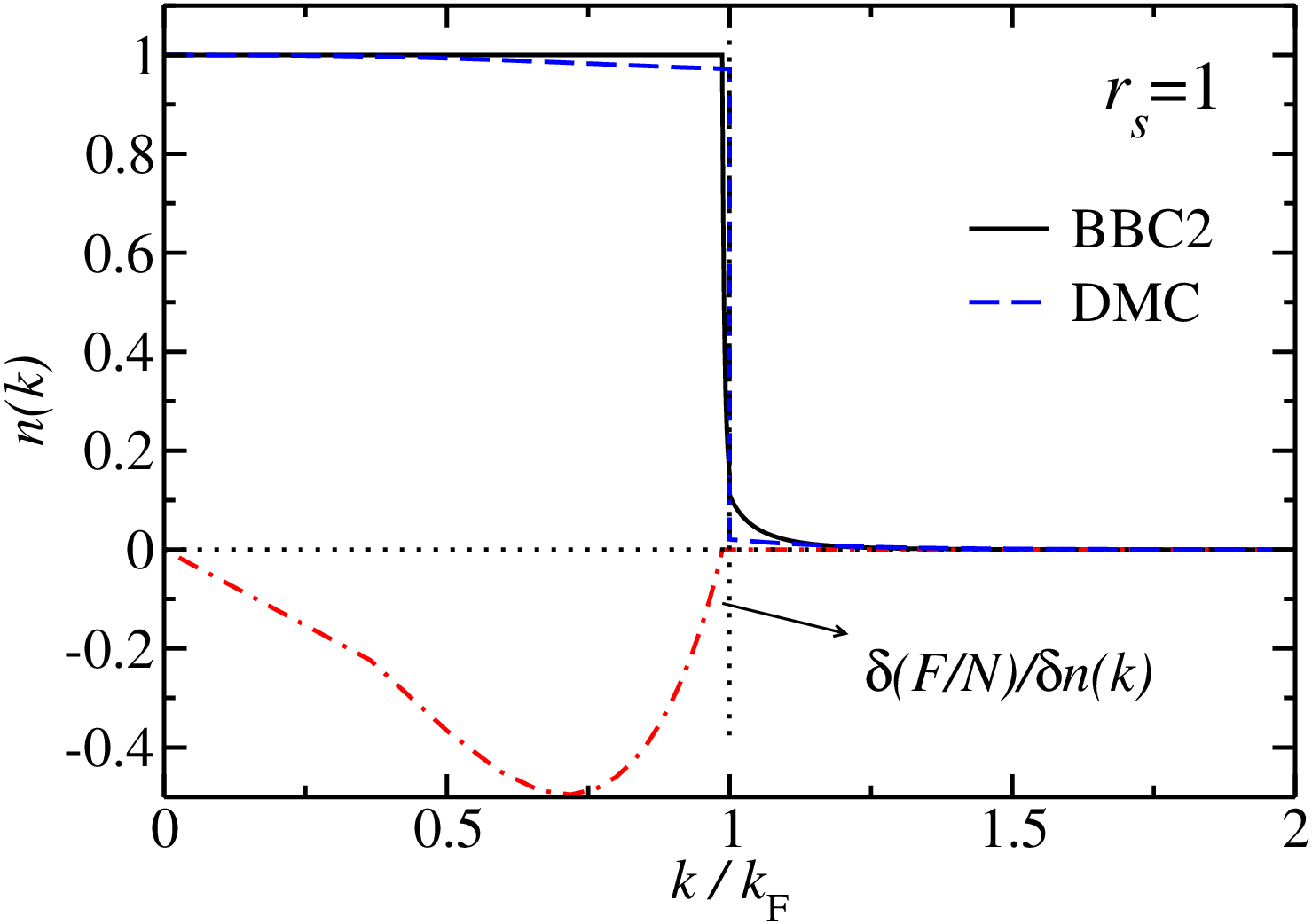}&  \ \ \ \ \ 
\includegraphics[angle=0,width=0.45\textwidth,clip]{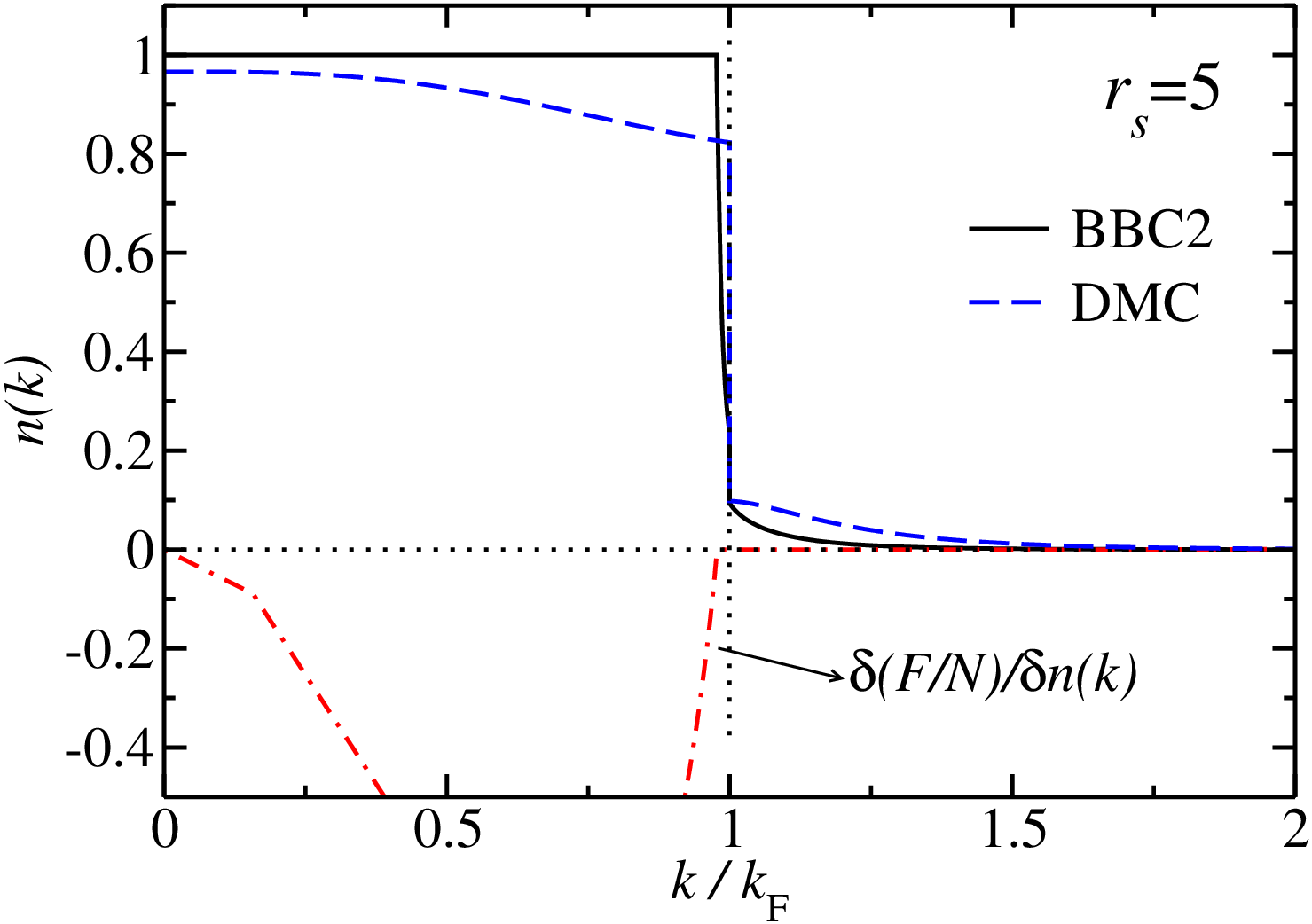}\\
\end{tabular}
\caption{\label{fig:momdistr} (Color online) The momentum distribution $n(k)$ of the HEG for $r_s$=1 (left)
and $r_s=5$ (right) 
calculated with the M\"uller functional, BBC1, and BBC2. 
BBC1 and BBC2 show a discontinuity of the momentum distribution at $k_{\rm F}$. For
comparison we include the fit to the DMC data
of Ortiz-Ballone~\cite{exactmom2}.  The derivative 
$\delta (F/N) / \delta n(k) $ is also plotted. 
Note that for small $k$ the derivative is not zero and  $n(k)$ is pinned at one.
}
\end{figure*}

\begin{figure}
\includegraphics[angle=0,width=0.9\columnwidth,clip]{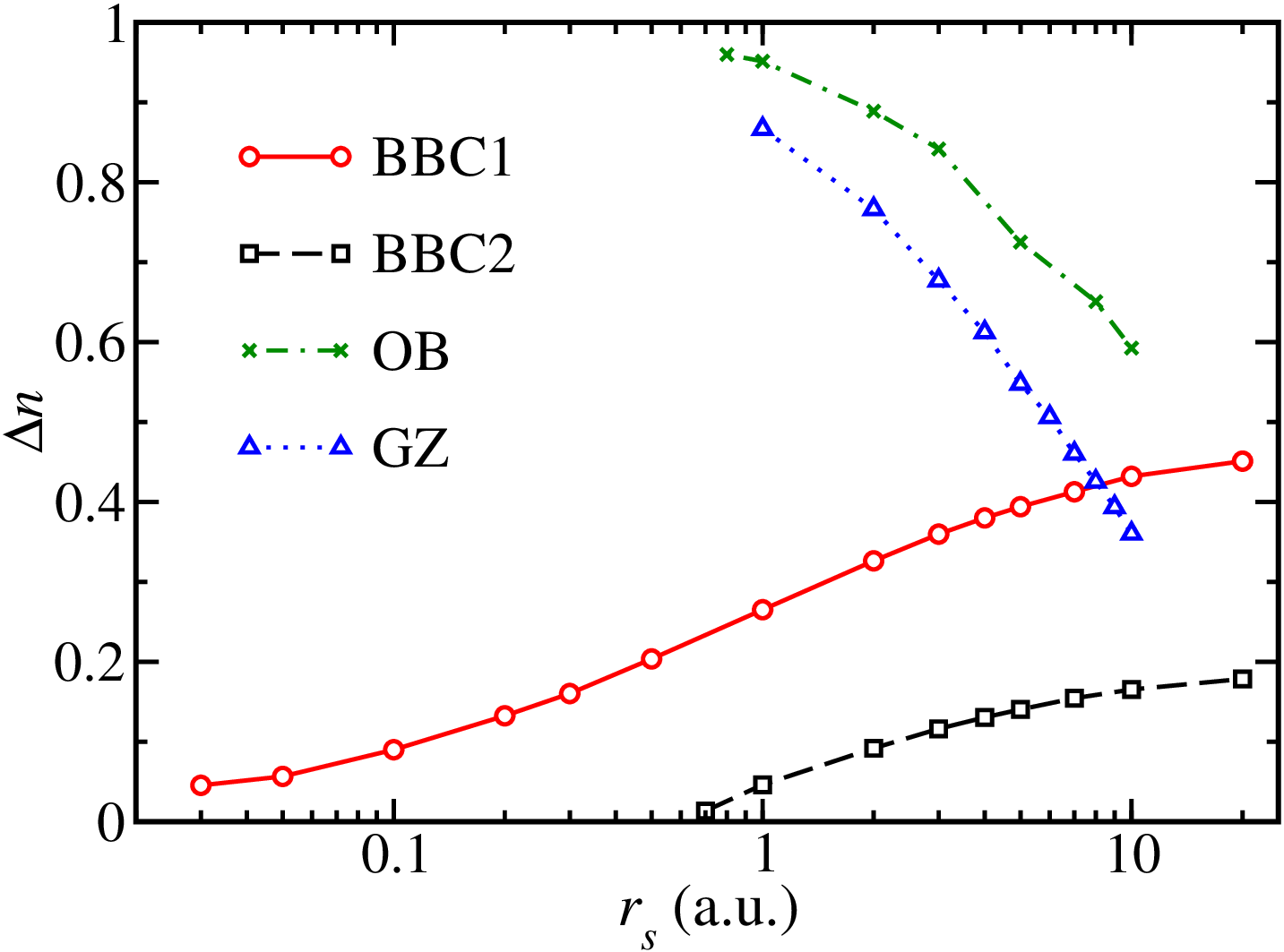}
\caption{\label{fig:zeta1} (Color online) The discontinuity $\Delta n$ of the momentum distribution at $k_{\rm F}$ for BBC1 and
BBC2 as a function of $r_s$, compared to the fits of Ortiz and Ballone (OB)~\cite{exactmom2}, and
Gori-Giorgi and Ziesche (GZ)~\cite{gori} to the DMC data.}
\end{figure}

A feature of the exact momentum distribution, namely the discontinuity at the 
Fermi wavevector $k_{\rm F}$,  is reproduced by the BBC functionals. 
The discontinuity is more pronounced for the BBC1 functional, as can be seen in 
Fig.~\ref{fig:momdistr},
where we plotted the momentum distribution of the HEG with $r_s=1$ and $r_s=5$  
using the M\"uller functional as well as BBC1 and BBC2. Contrary to BBC1 and BBC2 
the M\"uller functional does not yield a discontinuity. To our knowledge,
there is no report of any other 1-RDM functional reproducing this feature of the HEG.
To extract the size of the discontinuity quantitatively, 
we used two energy mesh points very close to $\epsilon_{\rm F}$ 
(at a distance of $\pm 10^{-8}\epsilon_{\rm F}$).
In Fig.~\ref{fig:zeta1}, we plot the size of the 
discontinuity $\Delta n$ as a function of $r_s$. As we see, it increases monotonically 
with $r_s$ and has the tendency to saturate for large $r_s$ for both BBC1 and BBC2. 
For BBC2 the discontinuity is substantially smaller than for BBC1 and it goes to zero 
at $0.6 < r_s < 0.7$. 
Both the size and the dependence on $r_s$ are in complete disagreement
with the exact theory, where $\Delta n$ is substantially bigger and decreases with $r_s$
as one can see from the two fits to the DMC data~\cite{exactmom2,gori}.

\begin{figure}
\includegraphics[angle=0,width=0.9\columnwidth,clip]{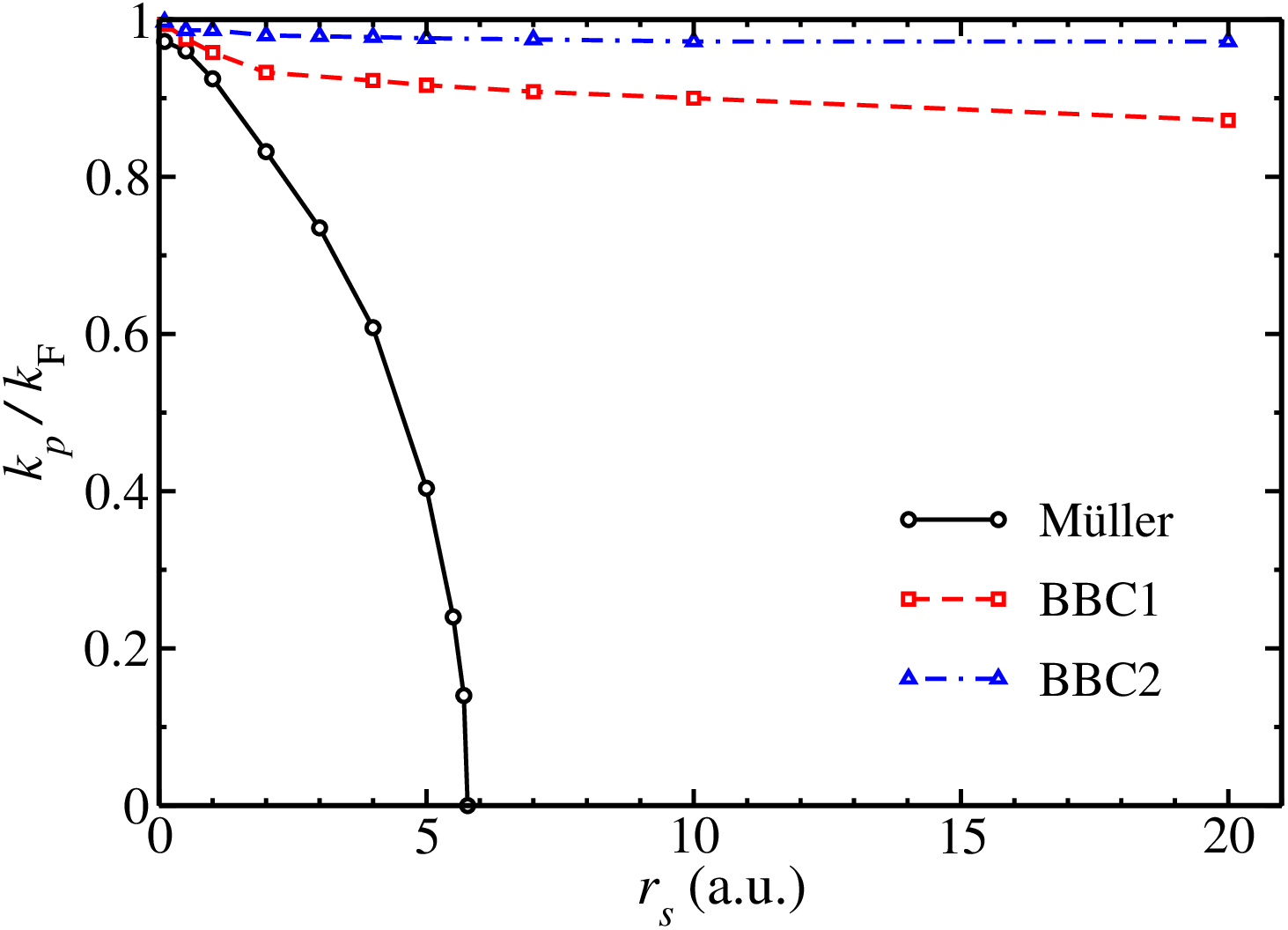}
\caption{\label{fig:ep_rs} (Color online) The wavevector $k_p$, where the optimal
momentum distribution becomes fractional, as a function of $r_s$.
 }
\end{figure}

We now turn to the question of state pinning. As we see in Fig.~\ref{fig:momdistr}, state
pinning is a common feature of all the functionals we employed. 
To verify that the states are truly pinned, we plot the functional derivative 
$\delta F /\delta n(k)$ which is non-zero for pinned states. As one expects,
the number of pinned states decreases with increasing $r_s$.
We define a wavevector $k_p$ below 
which the corresponding states are pinned, i.e. $n(k)=1$ for $k < k_p$.
In Fig.~\ref{fig:ep_rs}, we plot $k_p$
as a function of $r_s$ for the M\"uller and the two BBC functionals.
For all three functionals, $k_p$ decreases monotonically with $r_s$. 
For the M\"uller functional our numerical calculation confirms the analytic 
result~\cite{ciospernal} that above a critical value of $r_s=5.77$ there are
no pinned states, and therefore $k_p$ goes to zero at this value. 
This can already be seen in Fig.~\ref{fig:momdistr}, where, for the M\"uller functional 
at $r_s=5$, the derivative is very close to zero even for small wavevectors.
Interestingly, for BBC1 and BBC2 we found no such critical value up to
$r_s=20$. Indeed, for both BBC1
and BBC2 the decrease of $k_p$ is much smaller than for the M\"uller functional,
$k_p$ being almost constant for BBC2.

As we have seen, the performance of the BBC1,2 functionals for the HEG is 
improved significantly compared to previous functionals. This is especially 
remarkable, given that they were originally constructed to describe the 
dissociation of small molecules. 

\subsection{Improved functionals for the HEG}

In this section, we attempt to improve over the BBC functionals
for the HEG. We present two functionals which are simple modifications of the 
BBC1 functional. In both cases, the BBC1 functional is generalized 
by introducing an additional function of $r_s$
such that the correlation energy of the HEG is reproduced exactly for each $r_s$.
As exact results we regard the Perdew-Wang fit~\cite{PW} of the 
correlation energy obtained from diffusion Monte Carlo (DMC) calculations by 
Ceperley and Alder~\cite{exactmom1} and
Ortiz and Ballone~\cite{exactmom2}. The two Monte Carlo calculations yield almost identical 
correlation energies.

\begin{figure}
\begin{tabular}{cc}
\includegraphics[angle=0,width=0.45\columnwidth,clip]{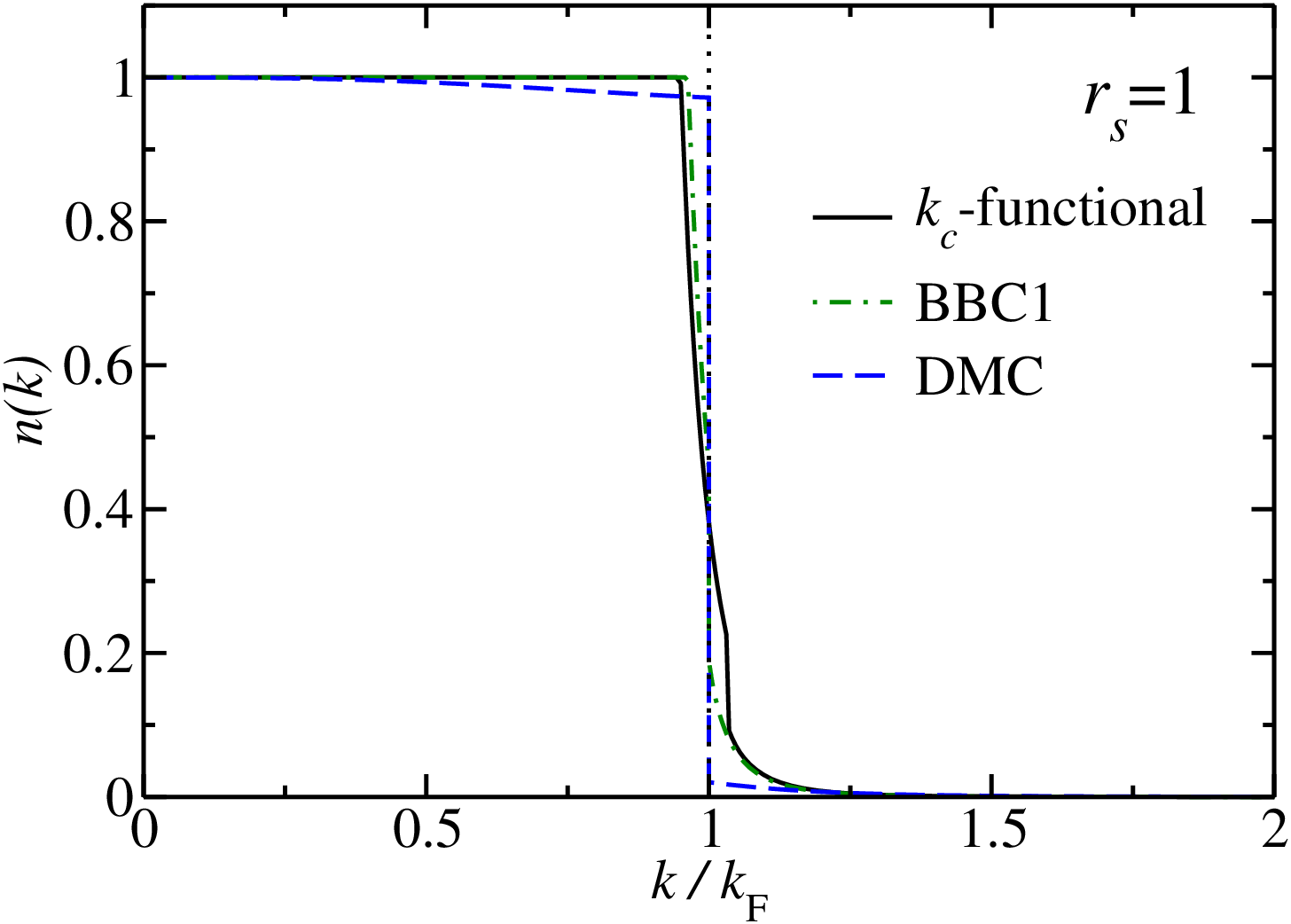}  \ \ \ & \ \ \ 
\includegraphics[angle=0,width=0.45\columnwidth,clip]{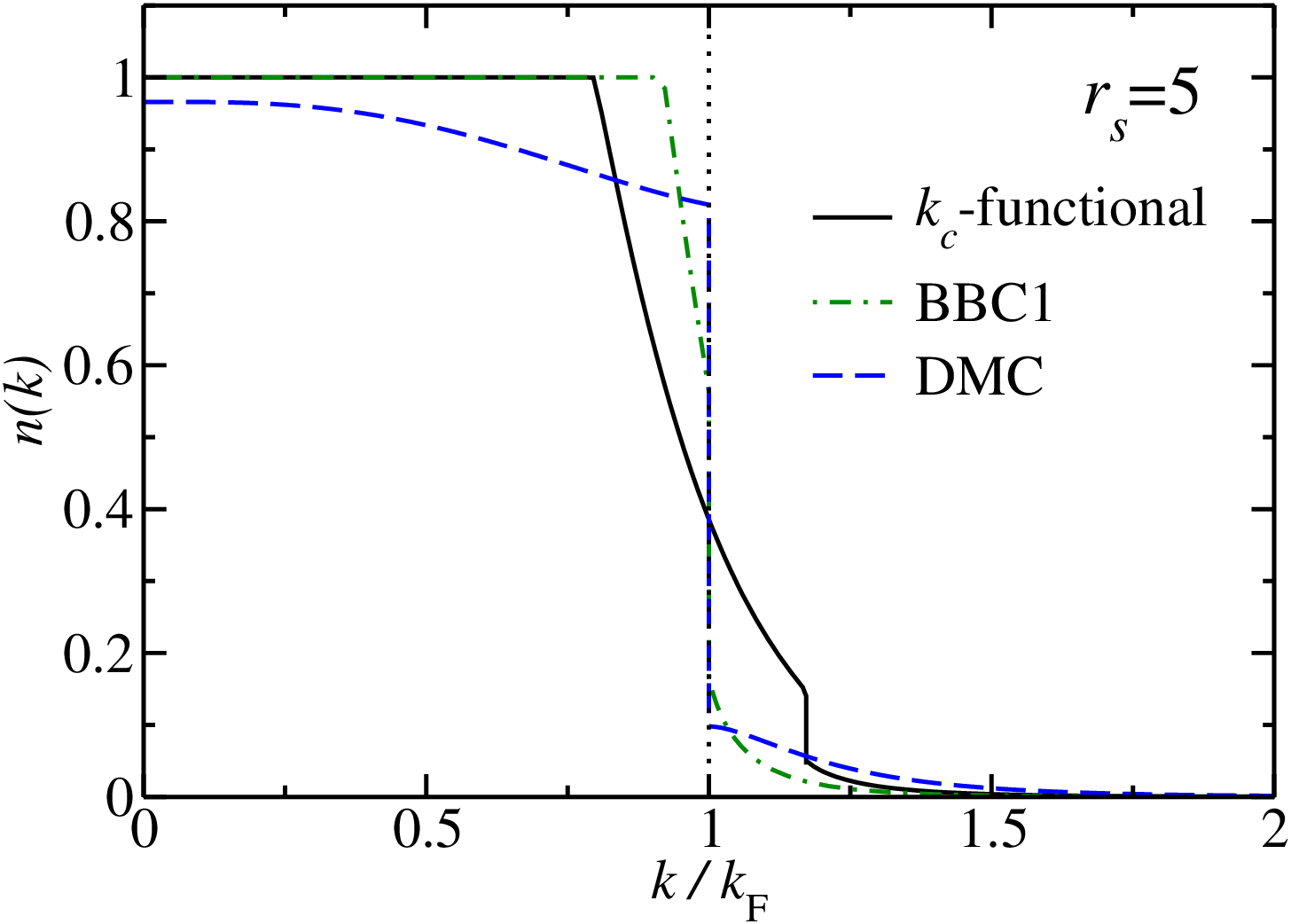} \\
\end{tabular}
\caption{\label{fig:ec_rs5} (Color online) The momentum distribution for the $k_c$-functional 
compared with the BBC1 and the fit to the DMC data of Ortiz-Ballone~\cite{exactmom2}, 
for  $r_s=1$ and $r_s=5$. For the $k_c$-functional the 
discontinuity is moved from $k_{\rm F}$ to $k_c$, i.e to $1.032k_{\rm F}$ and $1.172k_{\rm F}$
respectively.}
\end{figure}

For the first functional, we adjust the critical wavevector $k_c$, which 
is used to distinguish between strongly and weakly occupies states, instead
of using $k_c=k_{\rm F}$, as in BBC1. We call this functional the $k_c$-functional.
The corresponding function $f$ in Eq.~(\ref{eq:Exc}) then reads
\begin{equation}
 f(n({k_1}), n({k_2})) =  \left\{
 \begin{array}{rl}-\sqrt{n({k_1})\: n({k_2})}\,, & \quad \mbox{$k_1,k_2>k_c(r_s)$,} \\
                   \sqrt{n({k_1})\: n({k_2})}\,, & \quad \mbox{otherwise.} \\
 \end{array} \right.
\label{eq:ourfunct_0}
\end{equation}
We perform  the fitting of $k_c$ 
over the range of metallic densities, $0.5 \leq r_s \leq 5$. 
The results are compiled in Table~\ref{tab:ec}. For $r_s = 0.5$, $k_c\approx k_{\rm F}$, since this
point is almost exactly reproduced by BBC1 (see Fig.~\ref{fig:correl}). For $r_s > 0.5$, $k_c$ is a monotonically increasing
function of $r_s$.  Fitting $k_c$ has a strong impact on the momentum distribution which is 
displayed in Fig.~\ref{fig:ec_rs5}, for $r_s=1$ and $r_s=5$. 
It is not surprising that the discontinuity is 
displaced from $k_{\rm F}$ to $k_c$. Additionally, its size is reduced significantly compared
to BBC1. Both the displacement of the discontinuity and the decrease in the
step size are in disagreement with the exact result.

\begin{table}[t]
\begin{tabular}{c||c|c|c|c|c|c}
\hline\hline
$r_s$ & 0.5 & 1.0 & 2.0 & 3.0 & 4.0 & 5.0 \\ \hline
$k_c$/$k_{\rm F}$ & 0.994 & 1.032 & 1.085 & 1.122 & 1.155 & 1.172 \\
\hline\hline
\end{tabular}
\caption{\label{tab:ec} The the critical wavevector $k_c$ for some metallic 
densities. }
\end{table}

\begin{figure}
\includegraphics[angle=0,width=0.7\columnwidth,clip]{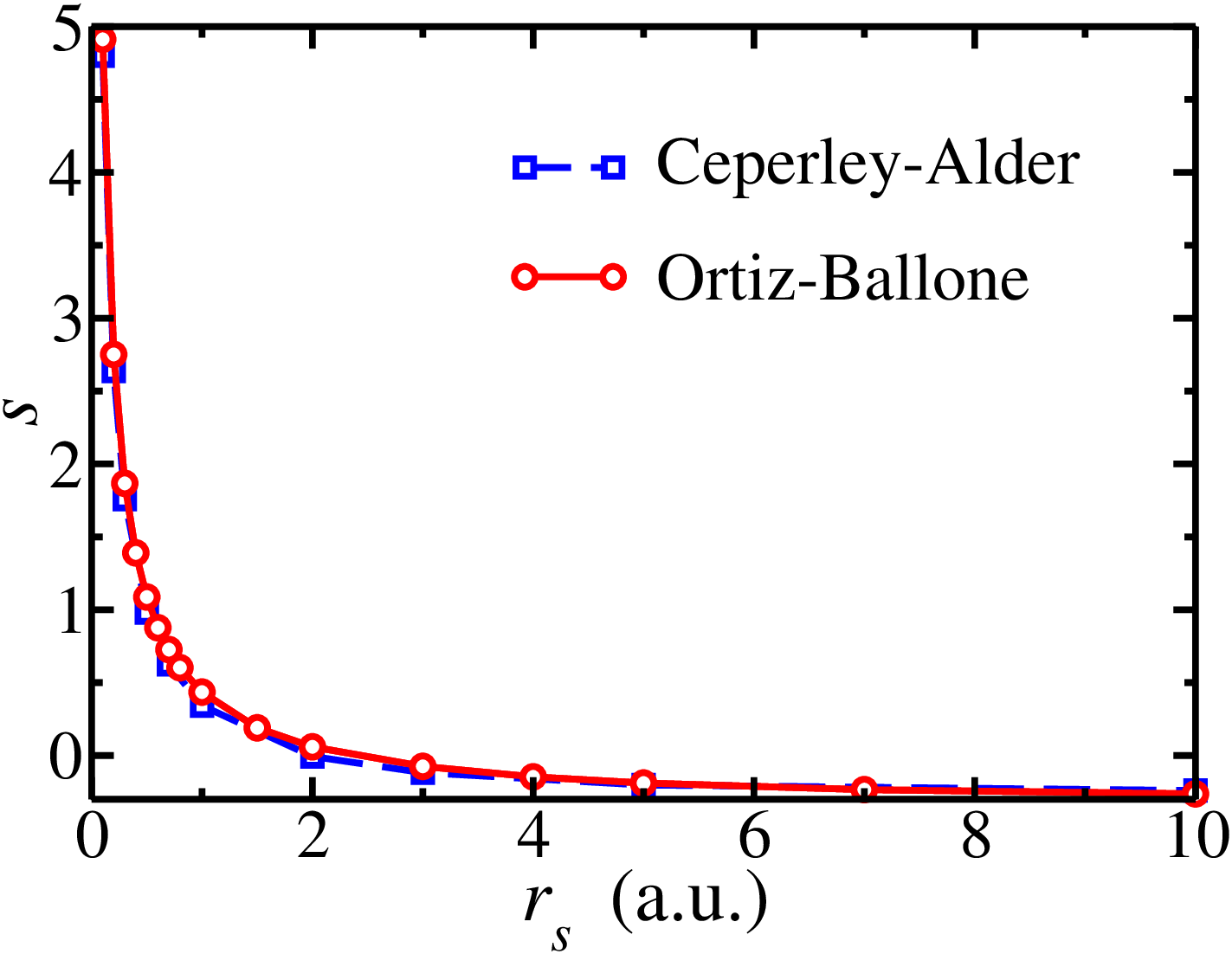}
\caption{\label{fig:s_rs} (Color online) The dependence of the fitting parameter 
$s$ on $r_s$ resulting
from fits to the Perdew-Wang parameterization~\cite{PW} of the correlation energy from
two different sets of DMC results:
the Ceperley and Alder~\cite{exactmom1} and the Ortiz and Ballone~\cite{exactmom2}.} 
\end{figure}

An alternative idea is to keep $k_c=k_{\rm F}$ fixed and consider a fitting parameter 
$s$ multiplying the exchange-like terms when both states, $\bk_1$ and $\bk_2$, are weakly occupied, i.e.
\begin{equation}
 f(n({k_1}), n({k_2})) =  \left\{
 \begin{array}{rl}-s(r_s)\,\sqrt{n({k_1})\: n({k_2})}\,, & \quad \mbox{$k_1,k_2>k_{\rm F}$,} \\
                   \sqrt{n({k_1})\: n({k_2})}\,, & \quad \mbox{otherwise.} \\
 \end{array} \right.
\label{eq:ourfunct}
\end{equation}
In this way, the parameter $s$ which is a kind of strength of the 
xc terms becomes a function of $r_s$, as indicated 
in Eq.~(\ref{eq:ourfunct}). We call this functional the $s$-functional.
 The values of $s$ for the fitting to the  Ortiz and 
Ballone DMC results are included in Table~\ref{table:s_functional}.
As we see, $s$ varies between $4$ and $-0.26$ 
for the range of densities, $0.1 \leq r_s \leq 10$, we considered.
In Fig.~\ref{fig:s_rs}, we show the resulting, monotonically decreasing
function $s(r_s)$.

\begin{table}
\setlength{\tabcolsep}{0.6truecm}
\begin{tabular}{cc}
\hline\hline
\begin{tabular}{cc}
$r_s$ &  s \\ \hline
0.1 & 4.913 \\
0.2 & 2.751 \\
0.3 & 1.867 \\
0.4 & 1.390 \\
0.5 & 1.087 \\
0.6 & 0.877 \\ 
0.7 & 0.727 \\
0.8 & 0.602 \\ 
\end{tabular} & 
\begin{tabular}{cc}
$r_s$ & s \\ \hline
1.0 & 0.435 \\
1.5 & 0.190 \\
2.0 & 0.059 \\
3.0 & -0.074 \\
4.0 & -0.146 \\
5.0 & -0.189 \\
7.0 & -0.234 \\
10.0 & -0.263 \\ 
\end{tabular} \\ \hline\hline
\end{tabular}
\caption{\label{table:s_functional}The fitted values of $s$ for various values of $r_s$ for the 
$s$-functional. $s$ was fitted to reproduce the correlation energies of the DMC calculation 
of Ortiz and Ballone~\cite{exactmom2}.}
\end{table}

\begin{figure}
\begin{tabular}{cc}
\includegraphics[angle=0,width=0.45\columnwidth,clip]{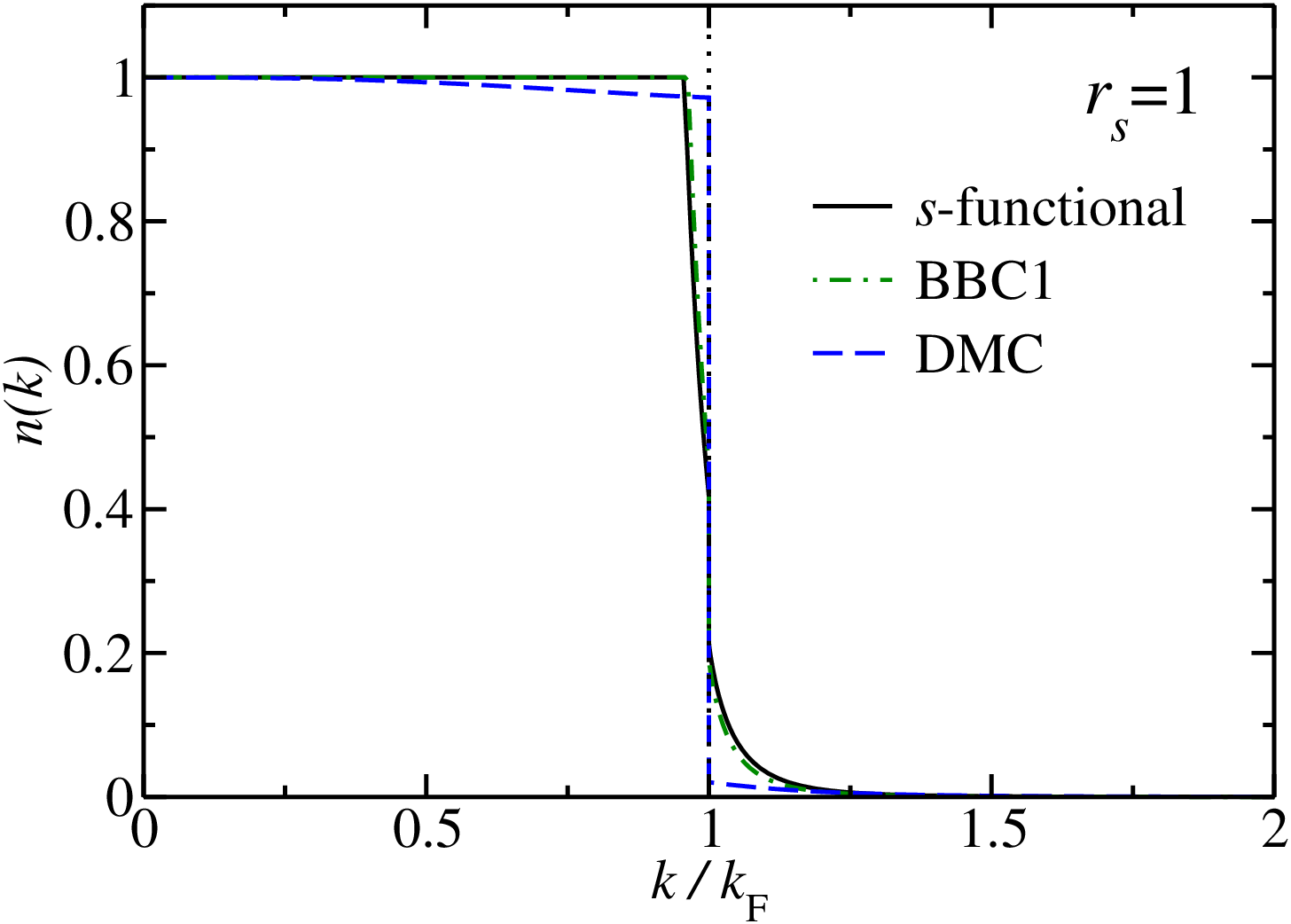} \ \ \ & \ \ \ 
\includegraphics[angle=0,width=0.45\columnwidth,clip]{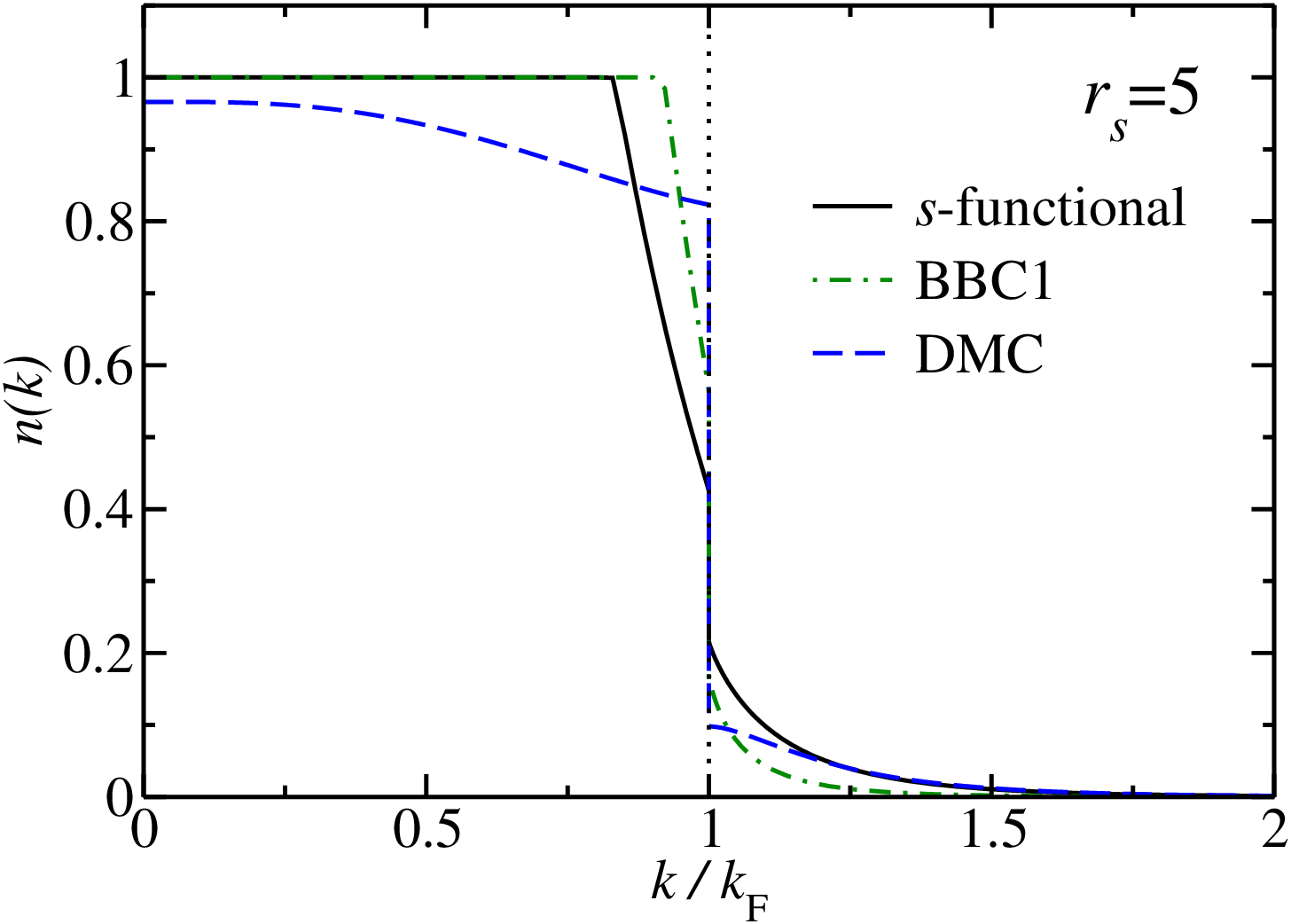} \\
\end{tabular}
\caption{\label{fig:n_k_fitts} (Color online) The momentum distribution given by the $s$-functional,
compared with the BBC1 and the fit to the DMC data of Ortiz-Ballone~\cite{exactmom2},
for $r_s=1$ ($s=0.435$), and $r_s=5$ ($s=-0.189$).  The discontinuity remains at 
$k_{\rm F}$ and is approximately equal to 0.2.}
\end{figure}

From Fig.~\ref{fig:n_k_fitts} one can see that one of the advantages of fitting $s$ instead of $k_c$
is that the discontinuity of the momentum distribution remains fixed at $k_{\rm F}$. 
In addition, its size is almost constant ($\approx 0.2$) as a function of $r_s$.
As we see in Fig.~\ref{fig:zeta1}, the exact discontinuity is significantly higher and it is a decreasing
function of $r_s$. Therefore, concerning the size of the discontinuity, the $s$-functional 
does not improve over the BBC1, which,
in the dilute gas limit, is close to the exact result.
However, the increasing behavior of BBC1  is improved by the $s$-functional which yields 
discontinuities that are almost constant as a function of $r_s$.

In conclusion, the $s$-functional results in momentum distributions that resemble 
the exact ones over the whole range of $r_s$ more closely than any of the other functionals
considered here.

\section{conclusion\label{sec:conclusion}}
We have applied a variety of 1-RDM functionals to the  
HEG. We show that the BBC functionals~\cite{gritsenko} yield a significant 
improvement over previous functionals as far as the correlation energy is concerned. In
addition, they yield a discontinuity of the momentum distribution at
the Fermi wavevector in resemblance of the exact HEG theory. However,
the size and the dependence on the density of this discontinuity are not 
in agreement with the quantum Monte Carlo results. 

By introducing an appropriately fitted function of $r_s$ in the BBC1 functional, we
demonstrate that the exact correlation energy of the HEG can be reproduced with a
smooth and monotonic fitting function.
For this function, we either use the critical wavevector $k_c(r_s)$ which
distinguishes between the strongly and weakly occupied states, or 
a strength $s(r_s)$ multiplying the exchange-like terms for two weakly occupied states. 
Both of these procedures were applied to the BBC1 functional. The two functionals,
yielding by construction the exact correlation energy of the HEG, are assessed by the quality
of the resulting momentum distributions. We show that choosing the second procedure, i.e. the 
$s$-functional, is superior to fitting $k_c$. The discontinuity $\Delta n$ of the momentum distribution 
resulting from the $s$-functional is nearly constant as a function of $r_s$ and hence
represents a significant improvement over BBC1 and BBC2.
However, the momentum distribution obtained by the $s$-functional still deviates significantly
from the exact one. To remedy this, more complicated strategies have to be considered, possibly 
with the introduction of more fitting parameters.

Our functional of choice, being derived from the HEG, is expected to yield 
good results for metallic systems. The application to finite as well as 
non-metallic periodic systems is not straightforward because $r_s$ is not well-defined
in these cases. Hence, the necessity arises to map $r_s$ onto other quantities characterizing 
these systems, or to involve a LDA-type prescription relating $s(r_s)$ to the local density.

\begin{acknowledgements}
This work was supported in part by the  Deutsche Forschungsgemeischaft within the program SPP 1145,
by the EXCITING Research and Training Network, and by the 
EU's Sixth Framework Program through the Nanoquanta Network of Excellence (NMP4-CT-2004-500198).
\end{acknowledgements}


%
%
%
%
%
%
\end{document}